\documentclass[3p,times,twocolumn]{elsarticle}

\usepackage{ecrc}

\volume{00}

\firstpage{1}

\journalname{Nuclear Physics B Proceedings Supplement}

\runauth{}

\jid{nuphbp}

\jnltitlelogo{Nuclear Physics B Proceedings Supplement}

\usepackage{amssymb}

\usepackage[figuresright]{rotating}

\usepackage{bbm}
\usepackage{footmisc}
\usepackage{hyperref}
\usepackage{natbib}
\usepackage{amsmath}
\usepackage{caption}
\usepackage{subcaption}
\usepackage{xspace}
\usepackage{slashed}

\newcommand{\TeV}{\ensuremath{\,\mathrm{TeV}}\xspace}

\newcommand{\ii}{\mathord{\mathrm{i}}}
\newcommand{\SM}{\ensuremath{\,\mathrm{SM}}\xspace}

\begin{document}

\begin{frontmatter}



\dochead{}

\title{%
\vspace*{-5cm}
\begin{flushright}%
{\scriptsize %
FTUV-14-1009, IFIC/14-63, KA-TP-27-2014, LPN14-118, SFB/CPP-14-77, TTK-14-27}
\end{flushright}
\vspace*{4.4cm}
CP-violating Higgs boson production in association with three jets via
  gluon fusion}



\author[a,b]{Francisco Campanario\fnref{1}}
\author[b,c]{Michael Kubocz}
\fntext[1]{Speaker. Prepared for the 37th International Conference on High Energy Physics (ICHEP 2014), 
2-9 Jul 2014, Valencia, Spain.}

\address[a]{Theory Division, IFIC, University of Valencia-CSIC, E-46980
  Paterna, Valencia, Spain}
\address[b]{Institute for Theoretical
  Physics, KIT, 76128 Karlsruhe, Germany.}
\address[c]{Institut f\"ur
  Theoretische Teilchenphysik und Kosmologie, RWTH Aachen
  University, D52056 Aachen, Germany}

\begin{abstract}
In these proceedings, we present results for Higgs production at the
LHC via gluon fusion with triple real emission corrections and
the validity range of the heavy-top effective theory approximation for
this process. For a general $\mathit{CP}$-violating Higgs boson, we
show that bottom-quark loop corrections in combination with large
values of $\tan \beta$ significantly distort differential
distributions.
\end{abstract}


\end{frontmatter}


\section{Introduction}
\label{sec:intro}
Higgs production in association with two jets via gluon fusion~(GF) is
a promising channel to measure the $\mathit{CP}$-properties of the
recently found new scalar boson~\cite{Aad:2012tfa,Chatrchyan:1471016}
with mass of 126~GeV.
The azimuthal angle correlations lead to a distinguishable
spectrum~\cite{DelDuca:2001eu,Odagiri:2002nd} sensitive to (1) the
$\mathit{CP}$-nature of the Higgs particle, shifting the positions of
the minima/maxima, (2) the Yukawa couplings, which change the
normalization of the cross sections, (3) parameters in theories beyond
the Standard Model~(BSM), like $\tan \beta$, which can significantly
distort phenomenologically important differential distributions. Higgs
production in association with two jets via GF at leading order~(LO)
is a loop induced process of order $\alpha_s^4$. The LO corrections
for a scalar Higgs particle were computed in
Ref.~\cite{DelDuca:2001eu} and for a general $\mathit{CP}$-violating
Higgs in Ref.~\cite{Campanario:2010mi}. Predictions within an
effective theory approximation are known at NLO
accuracy~\cite{Campbell:2006xx,vanDeurzen:2013rv}. They 
reduce the scale uncertainties considerably. However, the validity of
these predictions are restricted to Higgs/jet transverse momentum
and/or for Higgs masses lower than the top-quark mass. Furthermore,
this approximation provides no predictions for bottom-quark induced
loop contributions which can be relevant in BSM scenarios, e.g. due to a
strong enhancement via $\tan \beta=v_u/v_d$, the ratio of two
vacuum expectation values, appearing in two Higgs doublet
models~(2HDM), like the Minimal Supersymmetric Standard Model.

In these proceedings, we show results for the production of a general
$\mathit{CP}$-violating Higgs boson in association with three jets via
GF at LO beyond the heavy top-quark limit approximation, presented in
Ref.~\cite{Campanario:2014oua}. Results, for the $\mathit{CP}$-even
scalar case were given in
Ref.~\cite{Campanario:2013mga}~\footnote{Recent results within the
  effective theory approach are also available at NLO in
  Ref.~\cite{Cullen:2013saa} for a $\mathit{CP}$-even scalar Higgs
  boson and for the electroweak induced process in
  Ref.~\cite{Campanario:2013fsa}}. This process is an essential piece
to compute the full NLO corrections of Higgs plus two jet production
process via GF. Focusing particularly on the $gg\to ggg \Phi$
sub-process, being enhanced by large gluonic PDFs at the LHC, we study
the validity of the effective theory approximation and the possible
modifications of the azimuthal angle correlations due to the presence
of additional radiation. Additionally we investigate bottom-quark loop
contributions in combination with large values $\tan \beta$.

Sections~\ref{sect:the} and \ref{sect:setup} describe briefly the
theoretical background and details of the calculational set
up. Numerical results are presented in Sect.~\ref{sect:num} and
conclusions in Sect.~\ref{sect:con}.

\section{Theoretical Framework}
\label{sect:the}
A general $\mathit{CP}$-violating Higgs boson, $\Phi$, formed by a
mixing of a $\mathit{CP}$-odd, $A$, and a $\mathit{CP}$-even Higgs,
$H$, state,
\begin{equation} \Phi= H \cos \alpha +A \sin \alpha \,,
\end{equation}
can be described via the Lagrangian
\begin{equation}
{\cal L}_{\rm Yukawa}=\overline{q} \, ( y_q \cos \alpha+ \ii
\gamma_5 \tilde{y}_q \sin \alpha)\, q\, \Phi \, ,
\label{eq:lag} 
\end{equation}
where $y_q$ and $\tilde{y}_q$ denote the Yukawa couplings of the
$\mathit{CP}$-even and $\mathit{CP}$-odd components to fermions
$q$. In the SM, the smallness of the bottom-quark mass, $m_b$, leads
to a suppression of the corresponding Yukawa coupling,
$y_b^{\SM}=m_b/v$, with $v=246$~GeV, the SM vacuum expectation
value. Thus, corrections with bottom-quarks provide negligible
contributions in gluon fusion production modes. The remaining
top-quark loop effects can be therefore described within certain
limits in a simplified form by the heavy top-quark mass limit
approach. In this limit, top-quark loops are replaced by effective
$\text{Higgs}-\text{gluons}$ vertices, reducing the complexity of the
calculations considerably. The corresponding Lagrangian for a general
$\mathit{CP}$-violating Higgs is given by
\begin{equation} 
{\cal L}_{\rm eff} = \frac{\alpha_s}{12\pi m_t} \left( Y_t
G_{\mu\nu}^a\,G^{a\,\mu\nu} + \tilde{Y}_t \frac{3}{2} \,
G^{a}_{\mu\nu}\,\tilde{G}^{a\, \mu\nu}\right) \Phi \;,
\label{eq:ggphi}
\end{equation}
where $G^{a}_{\mu\nu}$ represents the gluon field strength and
$\tilde{G}^{a\, \mu\nu} = 1/2\,
G^{a}_{\rho\sigma}\,\varepsilon^{\mu\nu\rho\sigma}$ its dual. The
generalized Yukawa couplings $Y_t=\text{FF}_H \,y_t \,\cos \alpha $
and $\tilde{Y}_t=\text{FF}_A\, \tilde{y}_t\, \cos \alpha $ include
also form factors (FF$_{H,A}$). They are derived from the partial
decay width of the Higgs boson to two gluons and are therefore
proportional to triangle fermion-loops, re-introducing back missing
quark dependencies~\cite{Spira:1997dg,Djouadi:2005gj}. Their explicit
form can be found in Ref.~\cite{Campanario:2014oua}. %
As mentioned before, the validity of the heavy top-quark limit is
restricted to Higgs masses and values of the transverse momentum lower
than the top-quark mass. Additionally, in a 2HDM of type II, for large
values of $\tan \beta$, the Yukawa couplings with bottom-quarks
receive strong enhancement and therefore should be taken into
account. In the effective theory approach, these contribution are, of
course, absent. The up- and down-type Yukawa couplings for
the $\mathit{CP}$-odd components are given by,
\begin{align}
 \label{c:tb}
\tilde{y}^{\text{II}}_{u}= -\frac{\cot \beta}{v}m_u \qquad \text{and}
\qquad \tilde{y}^{\text{II}}_{d}= -\frac{\tan \beta}{v}m_d \; ,
\end{align}
where the SM-like vacuum expectation value $v=\sqrt{v_d^2
  +v_u^2}=246$~GeV and $\tan \beta = v_u/v_d$ are functions of the two
vacuum expectation values.
Henceforth, in our study of $\Phi jjj$ production, we will include
bottom-quark loop corrections to investigate their phenomenological
effects and test the validity bounds of the effective theory
approximation.

\section{Calculational set up}
\label{sect:setup}
In $\Phi jjj$ production, there are four contributing sub-processes,
\begin{align}  \label{3jsubproc}
q\, q \rightarrow q\, q\,g\, \Phi , \quad q\, Q \rightarrow q\, Q\,g\,
\Phi , \nonumber \\ q\, g \rightarrow q\, g\,g\, \Phi , \quad g\ g
\rightarrow g\, g\, g\,\Phi \, ,
\end{align}
together with the corresponding cross-related processes, which are not
written here explicitly. We show results for the purely gluonic
channel, which is enhanced by the large gluon flux at the LHC. This
sub-process is also the most complicated one, because it contains
massive hexagon one-loop diagrams, allowing us to test the numerical
stability of our program. Note that the LO contribution involves a $2
\to 4$ one-loop $\times$ one-loop calculation, thus, being subjected
to numerical instabilities of different
kinds~\cite{Campanario:2012bh}. The details of our calculation and the
performed checks can be found in
Refs.~\cite{Campanario:2013mga,Campanario:2014oua}. Here, we provide a
summary with the main features.
%
\begin{figure}[!ht]
\centering
\includegraphics[width=\columnwidth]{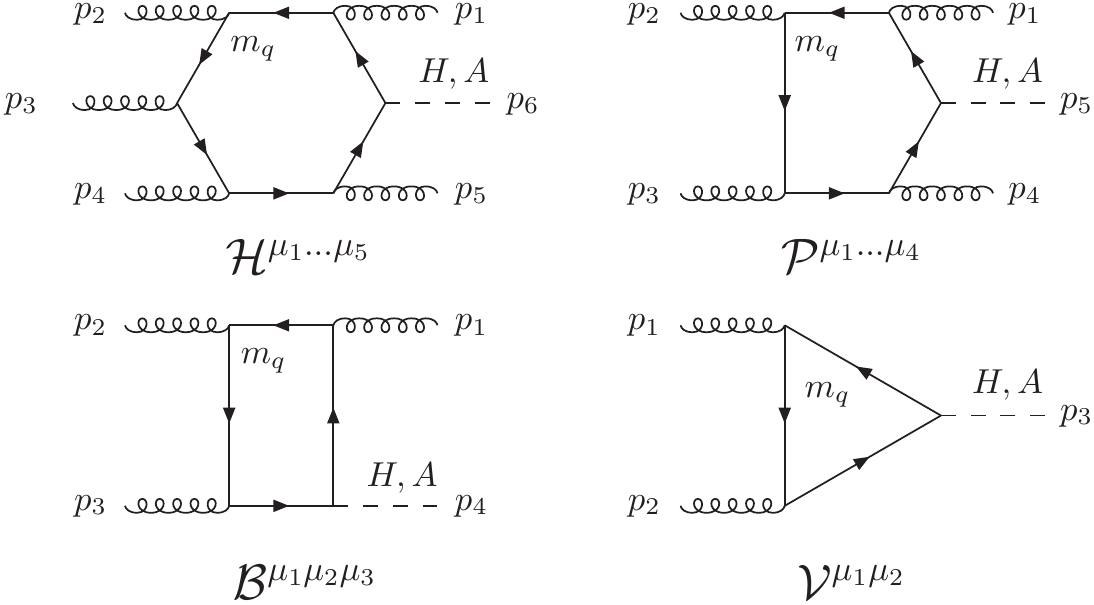}\quad
\vspace*{-1.8em}
\caption{Master Feynman diagrams}
\label{fig:diag}
\end{figure}
%
\begin{figure*}[!ht]
\centering
\includegraphics[width=0.95\columnwidth]{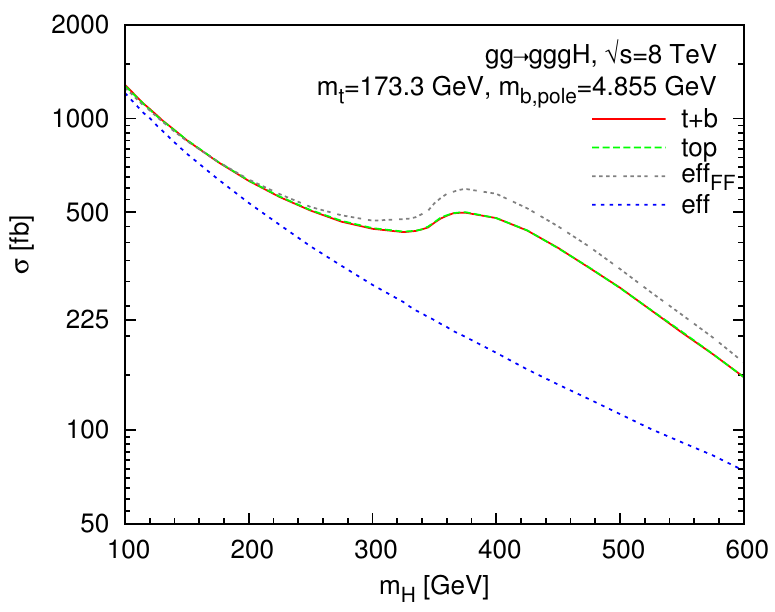}
\hfill
\includegraphics[width=0.95\columnwidth]{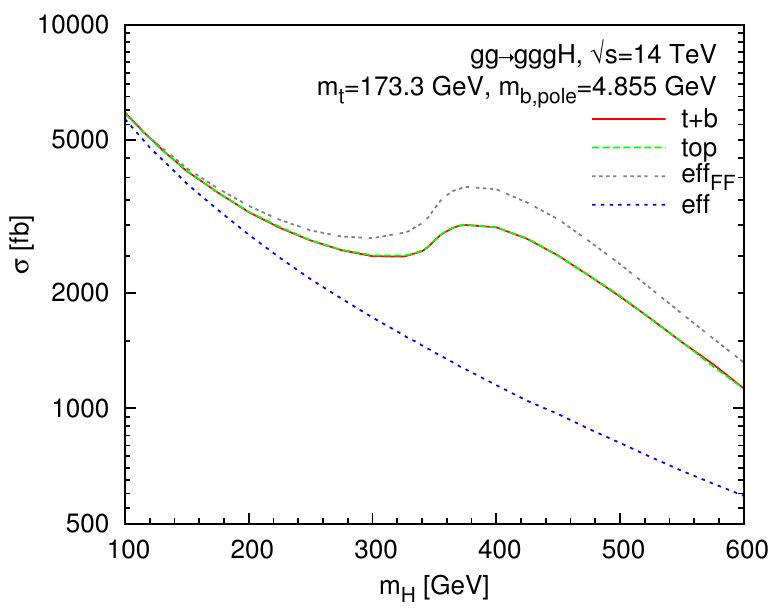}
\caption{Cross section of the $gg \rightarrow gggH$ scattering
  sub-process as a function of Higgs boson mass for c.m. energies of 8
  TeV and 14 TeV. Both panels show the effective (with and without FF)
  and the full theory contributions. The applied cuts are discussed in
  the text in more detail.}
\label{Hmass-scan}
\end{figure*}
Eight master Feynman diagrams~(Fig.~\ref{fig:diag}) involving four
$\mathit{CP}$-even and four $\mathit{CP}$-odd Higgs couplings to
fermions were computed with the in-house framework described in
Ref.~\cite{Campanario:2011cs} -- the attached gluons are considered to
be off-shell vector currents, which allow the addition of further
participating gluons. For this, we exploit the effective current
approach~\cite{Hagiwara:1985yu} to evaluate the loop
amplitudes. Furthermore, we apply Furry's theorem to reduce the number
of diagrams to be evaluated by a factor of two. The color factors were
computed by hand and cross-checked
with~\textsc{MadGraph4}~\cite{Alwall:2007st,Alwall:2011uj}. The
numerical stability is guaranteed with the help of Ward Identities
which are applied to every diagram (see Ref.~\cite{Campanario:2014oua}
for details).  In addition, we also generated results for this
sub-process using the effective theory approximation. With this
implementation, we can directly cross check our full theory
results. For $m_{t}=5\cdot 10^4$~GeV, the agreement is better than
one per ten thousand level. Finally, we have cross checked
successfully our results provided by the effective theory
approximation with the \textsc{MadGraph4} framework.

Both, the full and the effective theory implementation are available
via the GGFLO MC program, which is a part of the VBFNLO
framework~\cite{Arnold:2008rz,Arnold:2011wj,Arnold:2012xn}.
%
%
\begin{figure*}[!thb]
\centering
\includegraphics[width=0.95\columnwidth]{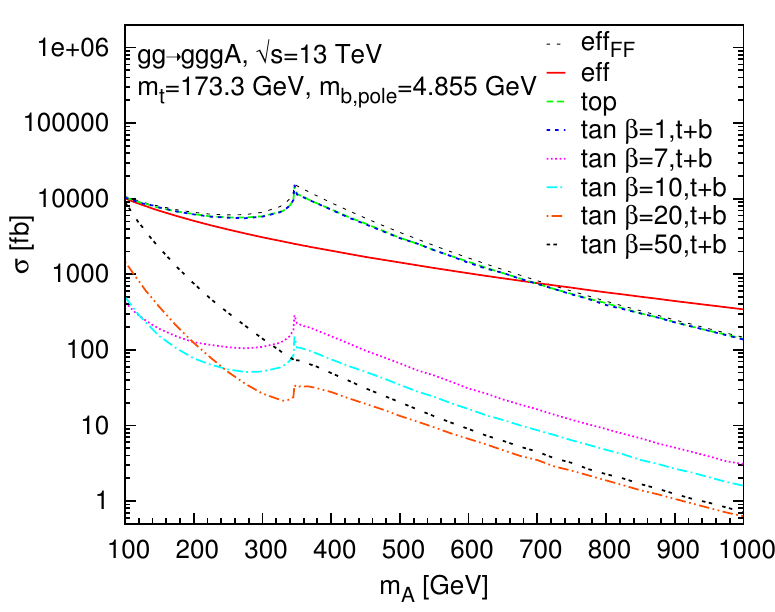}%
\hfill %
\includegraphics[width=0.95\columnwidth]{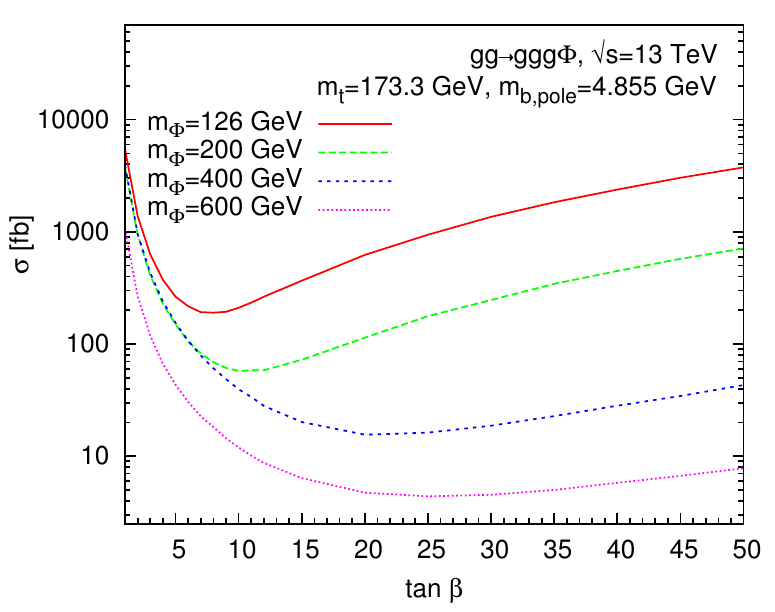}
\caption{Left: A + 3 jet cross section as a function of the
  pseudo-scalar Higgs boson mass $m_A$, for different values of $\tan
  \beta$. Right: $\Phi$ + 3 jet cross section as a function of $\tan
  \beta$ for several values of $m_{\Phi}$. The applied cuts are
  discussed in the text.}
\label{fig:massscanA}
\end{figure*}
\begin{figure*}[!thb]
\centering
\includegraphics[width=0.957\columnwidth]{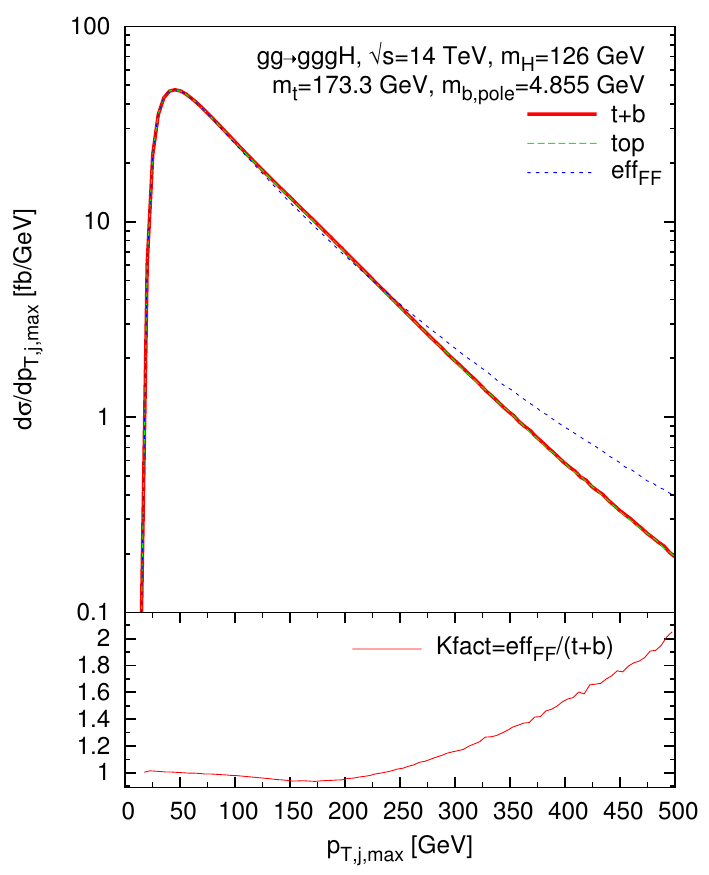}
\hfill
\includegraphics[width=0.957\columnwidth]{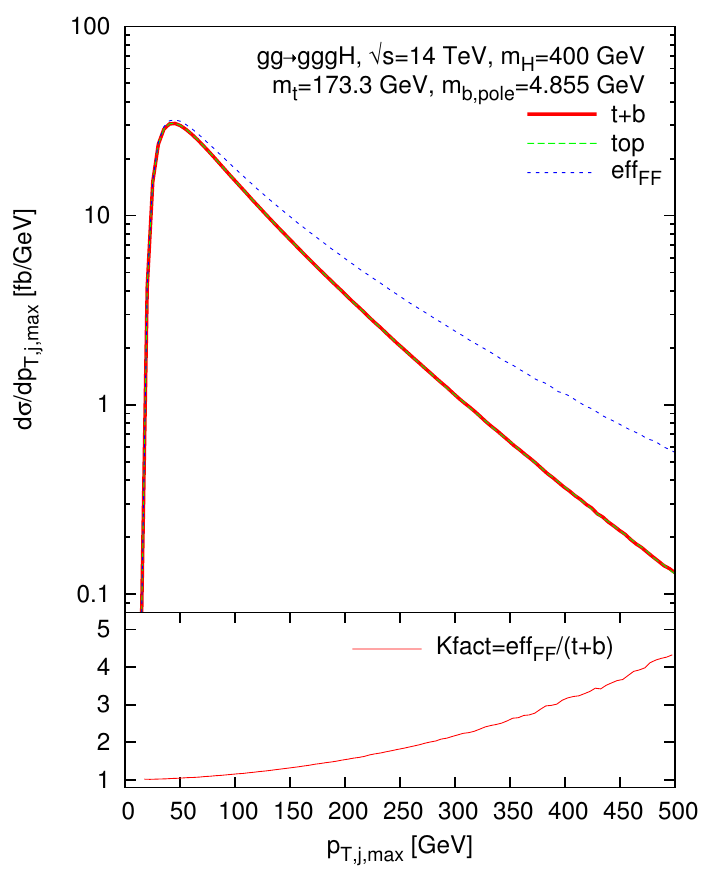}
\caption{Comparison of transverse-momentum distributions of the
  hardest jet of the $gg \rightarrow gggH$ scattering sub-process
  evaluated within the effective and loop-induced theory. Cuts are
  discussed in the text.}
\label{pthjet:diff}
\end{figure*}
\section{Numerical Results}
\label{sect:num}
In the following, results for integrated cross sections and
differential distributions are presented at the LHC for the
sub-process $gg\to ggg \Phi$.  To cluster jets, we use the
kT-algorithm and impose a minimal set of inclusive cuts to simulate
experimental acceptance capabilities. Jets are required to have a
$p_{T}^{j} > 20 \ \text{GeV}$ and to lie in the rapidity range
$|y_{j}| < 4.5$ with a cone radius of $R=0.6$. Additionally, jets are
ordered in terms of decreasing transverse momenta, $ p_{T}^{j_1} >
p_{T}^{j_2}> p_{T}^{j_3}$. As parton distribution functions~(PDFs), we
use CTEQ6L1~\cite{Pumplin:2002vw} with $\alpha_s(M_Z)=0.130$. The
Higgs boson is produced on-shell and without finite width effects. As
electro-weak input parameters, we choose $ M_Z = 91.188 \text{GeV}$,
$M_W=80.386 \text{GeV}$ and $G_F=1.16637\times 10^{-5}
\text{GeV}^{-2}$ and derive the weak mixing angle and the
electromagnetic coupling constant using SM tree level
relations. Except the top-quark with $m_t=173.3$~GeV and the
bottom-quark with a $\overline{\text{MS}}$~mass at
$\overline{m}_b(m_b)=4.2$~GeV, all other participating quarks are
taken to be massless. Additionally, following
Refs.~\cite{Spira:1997dg,Vermaseren:1997fq}, we use the relation
between the pole mass and the $\overline{\text{MS}}$ mass, within a
5-flavor scheme to take into account the evolution of $m_b$ up to the
reference scale $m_{\Phi}$. Note, that for the Higgs mass within the
range of 100-1000~GeV, the Yukawa couplings contain a 33-51$\%$
smaller $m_b$ than the pole mass of 4.855~GeV utilized in the loop
propagators.

\begin{figure*}[!thb]
\centering
\includegraphics[width=0.95\columnwidth]{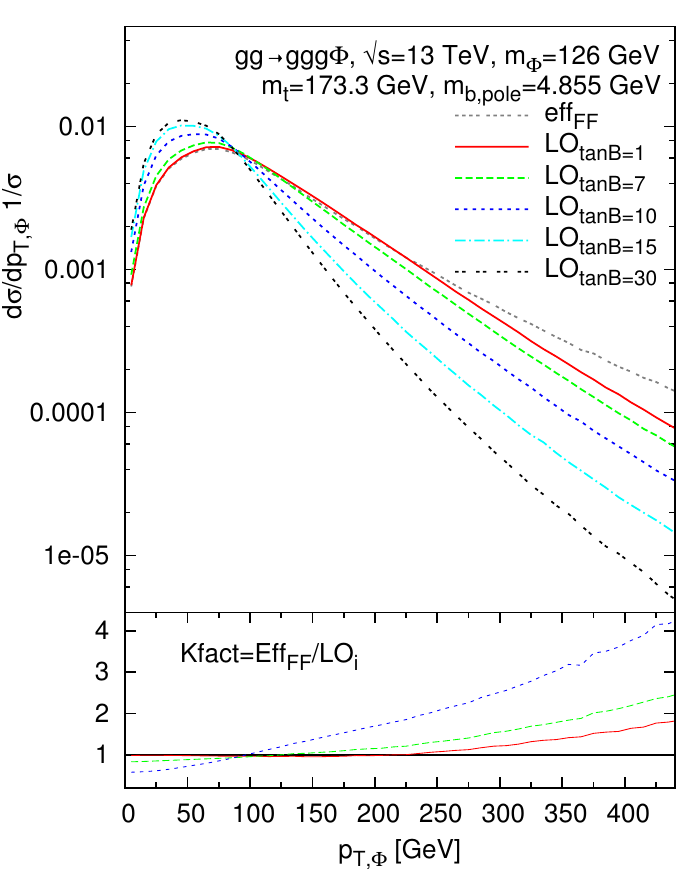}
\hfill
\includegraphics[width=0.950\columnwidth]{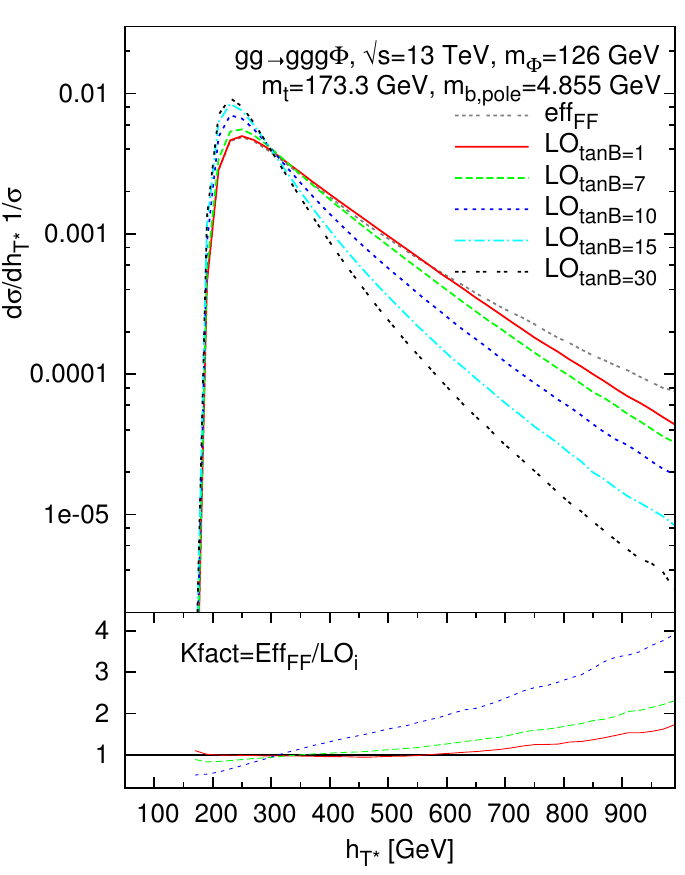}
\caption{The transverse-momentum distributions of the Higgs boson
  $\Phi$ (left panel) and the transverse scalar sum (right panel), are
  plotted. Details are described in the text.}
\label{ptAjet:diff}
\end{figure*}

The renormalization and factorization scales, are defined by
$\alpha_s^{5}(\mu_R) = \alpha_s(p_T^{j_1}) \alpha_s(p_T^{j_2})
\alpha_s(p_T^{j_3}) \alpha_s(p_{\Phi})^2$ and $\mu_F=(p_{T}^{j_1}
p_{T}^{j_2}p_{T}^{j_3})^{1/3}$.
In the following, we present results for three different scenarios:
\begin{eqnarray} \text{H~} (\text{CP-even}): & \alpha=\pi/2,&
y_q=y_q^{\text{SM}}\\ \text{A~} (\text{CP-odd}): & \alpha=0, & \tilde{y}_q=
\tilde{y}_q^{II}~~~~~~~~~\\ \Phi ~(\text{mixed state}): & \tan \alpha = 2/3,&
y_q = \tilde{y}_q= \tilde{y}_q^{II}.~~~
\label{eq:scenario}
\end{eqnarray} For the effective theory approximation, we show results with and
without form factors (FF). The total cross section as a function of
the Higgs boson mass for the $\mathit{CP}$-even Higgs boson is
presented in Fig.~\ref{Hmass-scan} at a center of mass energy (c.m.)
of $8 \TeV$ (left panel) and $14 \TeV$ (right panel). As expected,
bottom loop corrections are negligible within the SM. The effective
theory approximation provides accurate results ($10\%$) up to Higgs
masses of about 150 (175)~GeV at a c.m energy of 8~(14)~$\TeV$. The
application of form factors extends the validity range up to
approximately 300~GeV.  Additionally, it reproduces the threshold
enhancement at $m_H = 2m_t$. Beyond 300~GeV, the total cross section
is overestimated up to 20~(25)~$\%$ at 8~(14)~TeV c.m. energy for
$m_H=400$~GeV, and converges afterwards slowly to the full theory
result for the shown range of the Higgs mass.

For small values of $\tan \beta$, similar results are obtained for a
$\mathit{CP}$-odd Higgs boson (see the left panel of
Fig.~\ref{fig:massscanA}). For large values of $\tan \beta$, the bottom
loop corrections dominate and the effective theory is not valid
anymore. Note the fast decrease of the cross
section for bottom-quark loop dominating configurations (see the $\tan
\beta= 50$ curve). For the mixed state scenario (right panel), we show
the total cross section for $\Phi jjj$ production as a function of
$\tan \beta$ for different values of the Higgs mass. Note that the
minimum of the distribution, obtained for $\tilde{y}_t^{II} \approx
\tilde{y}_b^{II}$, is shifted to larger $\tan \beta$ values with
increasing Higgs mass. This is due to the fact that the bottom-quark
loop corrections decrease faster with increasing Higgs masses, hence,
one needs bigger values of $\tan \beta$ to set effectively both Yukawa
couplings to equal strengths.

Next, we show selected differential distributions. In
Fig.~\ref{pthjet:diff} the differential distribution of the
transverse momentum of the leading jet is presented for a light (left)
and heavy (right) Higgs boson. One can clearly see that even for light
Higgs bosons large differences appear between the effective theory,
including the form factors, and the full theory. The differences
between both approaches are illustrated with the help of a $K$-factor
defined as $\text{eff}_{\text{FF}}/(t+b)$, pictured in the lower
panels. As expected, the effective theory approximation yields
accurate results up to $p_{T}^{j\text{max}} < 200$~GeV. For a Higgs
mass of 400~GeV, where the largest discrepancy is observed (about
$25\%$ at the total integration level) between the full and the
effective theory including FFs, the differences are even more
prominent. The effective theory predicts harder emissions and overestimates
the full theory result up to a factor of 5.%

In Fig.~\ref{ptAjet:diff}, for the mixed state scenario and for small
values of $\tan \beta$, similar effects are observed in the
differential distributions of the transverse Higgs momenta~(left) and
the transverse scalar sum of the system~(right) defined as $h_{T^*}=
\sum _i p_T^{j_i} + \sqrt{p_{T,\Phi}^2+ m_{\Phi}^2}$. For $\tan
\beta=1$, despite the fact that the normalization is well predicted by
the effective theory, large differences up to a factor of 2
appear. These discrepancies become larger with increasing values of
$\tan \beta$ and additionally stronger softening effects are visible
resulting from bottom-quark loop corrections .
\begin{figure*}[!th]
\centering
\includegraphics[width=0.957\columnwidth]{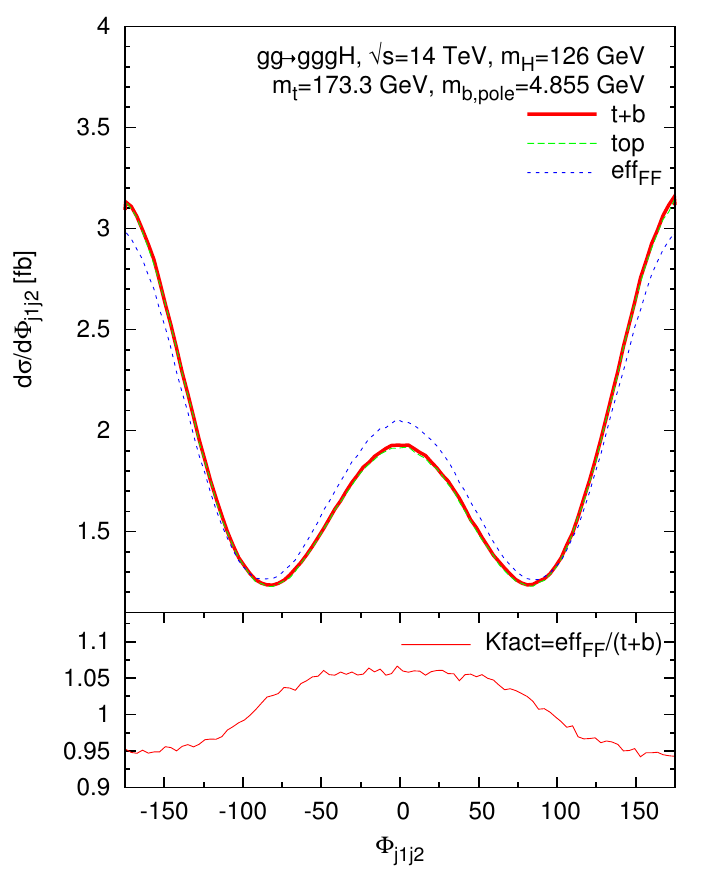}
\hfill
\includegraphics[width=0.957\columnwidth]{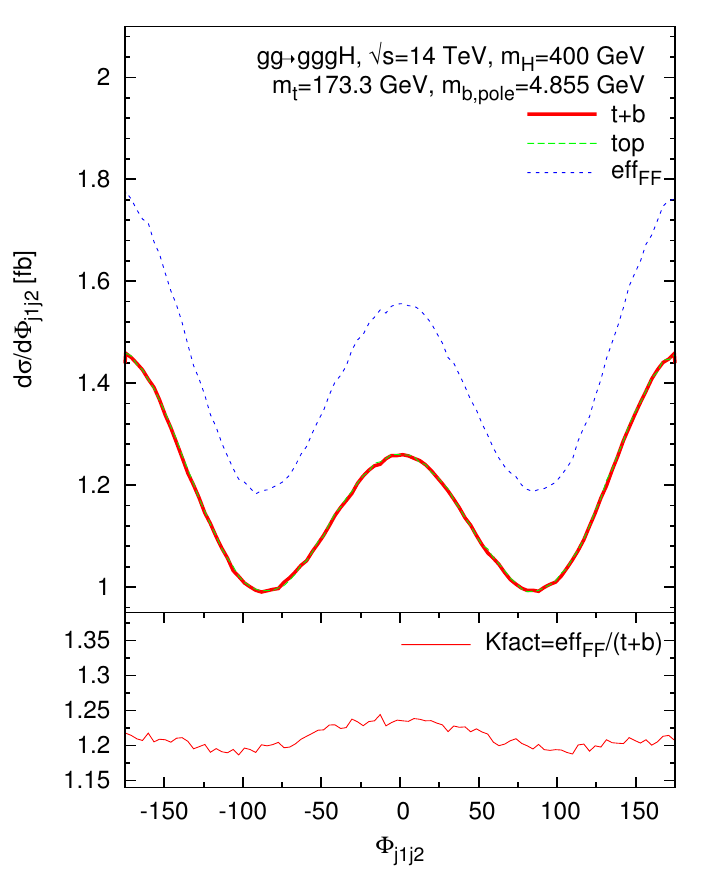}

\caption{Comparison of the azimuthal angle distributions of the $gg
  \rightarrow gggH$ scattering sub-process evaluated within the
  effective and loop-induced theory. Cuts are discussed in the text.}
\label{azim:diff}
\end{figure*}
\begin{figure}[!th]
\centering
\includegraphics[width=0.95\columnwidth]{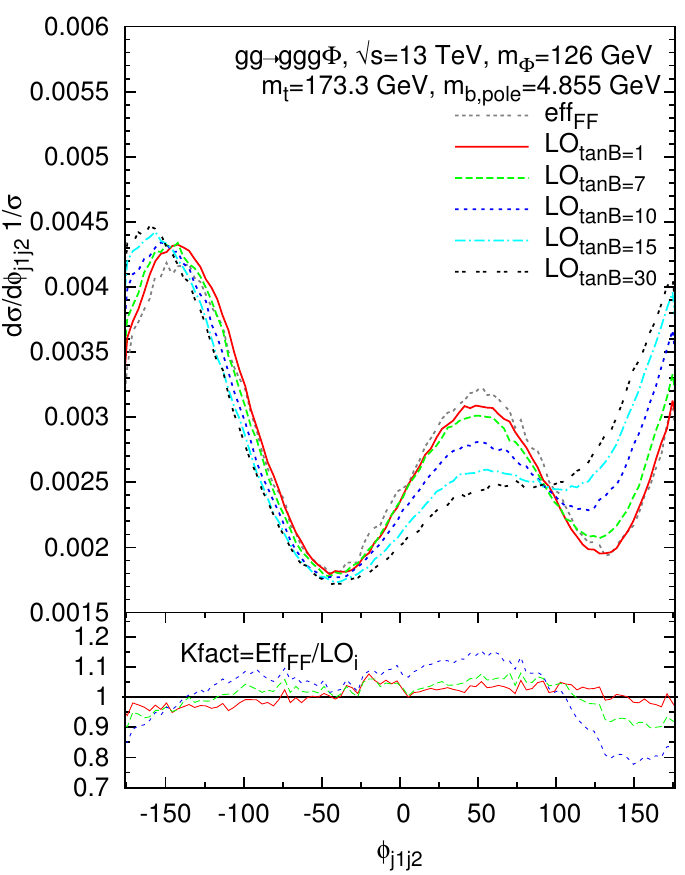}
\caption{Azimuthal angle correlation $\phi_{j_1j_2}$ of the two
  hardest jets. Cuts are discussed in the text.
}
\label{phi:diff}
\end{figure}

Next, we investigate the azimuthal angle distribution, defined as the
difference of the azimuthal angle between the more-forward and the
more-backward of the two tagging jets
($\phi_{j_1j_2}=\varphi_F-\varphi_B$). To increase the sensitivity of
the $\phi_{j_1j_2}$ distribution to the $\mathit{CP}$-structure of the
Higgs couplings, we slightly modify the applied cuts in the following
way
\begin{align} 
\label{ICphi} 
p_{T}^{j} > 30 \ \text{GeV},\;  \hspace{0.1cm} \Delta \eta_{j_1j_2}>3 .\;
\end{align}
In Fig.~\ref{azim:diff}, one can observe that the presence of
additional soft radiation does not distort the characteristic shape
obtained in the Higgs plus two jet production. In the left panel, for
a 126~GeV massive Higgs boson, the effective theory approach
approximates accurately the full theory with a small deviation of
5~$\%$. For a 400~GeV heavy Higgs boson, the shape is well reproduced,
but 20~\% off in the whole spectrum. The same distribution is plotted
in Fig.~\ref{phi:diff} for the $\mathit{CP}$-mixed state
scenario. Here, the shift of the minimum is given by the relation,
\begin{align} \tan \Delta \phi_{j_1j_2} = \frac{3}{2} \frac{\tilde{y}_q}{y_q }
\tan \alpha\,.
\end{align}
Hence, in our model the minima are shifted to
$\phi_{j_1j_2}^{\textnormal{min}}= -45^o,135^o$. It is clearly visible
that the presence of additional radiation does not alter the main
characteristics of the azimuthal angle correlations.  Additionally,
one observes that the effective theory approximation accurately
reproduces the shape of the $\phi_{j_1j_2}$ distribution. However, in
the full theory, it receives additionally kinematic distortions caused
by the balance of the transverse momenta of the jets with respect to
that of the Higgs boson due to conservation of momenta, and the softer
momentum spectrum of the jets and the Higgs boson for high values of
$\tan \beta$ (see Fig.~\ref {ptAjet:diff}).

\section{Conclusions}
\label{sect:con}
In these proceedings, we have presented results for the Higgs
production in association with three jets. We have focused on the
production process $gg\to ggg \Phi$, where $\Phi$ corresponds to a
general $\mathit{CP}$-violating Higgs boson, and showed results for
three different scenarios. %
We have studied the validity of the effective theory approximation and
confirmed that additional soft radiation does not alter the main
characteristics of the azimuthal angle correlations. %
Additionally, we have shown that bottom-quark loop contributions in
combination with large values of $\tan \beta$ can lead to visible
distortions of phenomenologically important differential distributions.

\section{Acknowledgments}
We thank Dieter Zeppenfeld for helpful discussions and collaborations at 
early stages of this project. 
This work was partially funded by the Deutsche Forschungsgemeinschaft
via the Sonderforschungsbereich/Transregio SFB/TR-9 Computational
Particle Physics. MK acknowledges support by the Grid Cluster of the
RWTH-Aachen.  FC acknowledges financial support by the IEF-Marie Curie
program (PIEF-GA-2011-298960) and partial funding by the LHCPhenonet
(PITN-GA-2010-264564) and by the MINECO (FPA2011-23596).




\nocite{*}
\bibliographystyle{elsarticle-num}
\biboptions{sort&compress}
\bibliography{ggA_p}






\end{document}